\documentclass[12pt]{article}

\renewcommand{\d}{{\rm d}}
\renewcommand{\O}{{\cal O}}
\newcommand{\overO}{\overline{\cal O}}
\newcommand{\overOm}{\overline{\Omega}}
\newcommand{\overT}{\overline{T}}
\newcommand{\overU}{\overline{U}}
\newcommand{\dotOm}{\dot{\Omega}}
\newcommand{\dotom}{\dot{\omega}}
\newcommand{\son}{\sigma^1}
\newcommand{\tson}{\tilde{\sigma}^1}

\begin{document}

\title{Discrete Faddeev action for the tetrad fields strongly varying along different coordinates
}

\author{V.M. Khatsymovsky \\
 {\em Budker Institute of Nuclear Physics} \\ {\em of Siberian Branch Russian Academy of Sciences} \\ {\em
 Novosibirsk,
 630090,
 Russia}
\\ {\em E-mail address: khatsym@gmail.com}}
\date{}
\maketitle
\begin{abstract}
Faddeev gravity using a $d$-dimensional tetrad (normally $d = 10$) is classically equivalent to general relativity (GR).

The discrete Faddeev gravity on the piecewise flat spacetime normally assumes slowly varying metric and tetrad from vertex to vertex. Meanwhile, Faddeev action is finite (although not unambiguously defined) for discontinuous tetrad fields thus allowing, in particular, to consider a surface as consisting of virtually independent elementary triangles, and its area spectrum as the sum of elementary area spectra. In the discrete connection form, area tensors are canonically conjugate to SO(10) connection matrices, and earlier we have found the elementary area spectrum, which is nonsingular just at large connection or the strongly varying fields constituting a kind of "antiferromagnetic" structure.

We appropriately define discrete {\it connection} Faddeev action to unambiguously determine the discrete Faddeev action for the strongly varying fields, but weakly varying metric, equivalent in the continuum limit to the GR action with this metric. Previously, we considered large variations in only one direction, now we use an ansatz in some respects less common, but overall, probably the most common.

A unified simplicial connection representation is written out (depending on an auxiliary connection) both for the discrete Faddeev action and for the Regge action.

\end{abstract}


MSC classes: 83C27; 53C05

keywords: general relativity; embedded gravity; Faddeev gravity; piecewise flat spacetime; Regge calculus; discrete connection

\section{Introduction}

Faddeev gravity can be considered as a theory in which the metric is composed of $d$ vector fields $f^A_\lambda$,
\begin{equation}                                                            
g_{\lambda \mu} = f^A_\lambda f_{\mu A}.
\end{equation}

\noindent Typically $d = 10$. After introducing a non-Riemannian connection $\tilde{\Gamma}^\lambda_{\mu \nu} = f^\lambda_A f^A_{\mu , \nu}$ and forming the Riemannian tensor for it, the Faddeev action $\int R \sqrt{g} $ $ \d^4 x$ can be written as
\begin{equation}\label{Faddeev}                                             
\int \left( f^\lambda_{A , \lambda } f^\mu_{A , \mu } - f^\lambda_{A , \mu } f^\mu_{A , \lambda } \right) \Pi^{AB} \sqrt{g} \d^4 x, ~~~ \Pi^{AB} = \delta^{AB} - f^{\lambda A} f^B_\lambda.
\end{equation}

\noindent The projector $\Pi^{AB}$ is called the {\it vertical} projector, and $\Pi^{AB}_{||} = \delta^{A B} - \Pi^{AB}$ is called the {\it horizontal} projector. Varying the action with the help of $\Pi_{A B} \delta / \delta f^\lambda_B$ gives the so-called vertical equations of motion,
\begin{equation}\label{vertical}                                            
b^\mu_{\mu A} T^\nu_{\nu \lambda} + b^\mu_{\lambda A} T^\nu_{\mu \nu} + b^\mu_{\nu A} T^\nu_{\lambda \mu} = 0,
\end{equation}

\noindent which can be considered as equations for the torsion $T^\lambda_{\mu \nu} = \tilde{\Gamma}^\lambda_{\mu\nu} - \tilde{\Gamma}^\lambda_{\nu \mu}$; $b^\lambda_{\mu A} = \Pi_{AB} f^{\lambda B}_{, \mu}$. For random fields $f^A_\lambda$, this system gives $T^\lambda_{\mu \nu} = 0$, and we return to the Hilbert-Einstein action; the horizontal equations turn out to be the Einstein equations.

One can also think of $f^A_\lambda$ as a tetrad, which is in tangential pseudo-Euclidean spaces embedded into ambient flat ten-dimensional pseudo-Eucli\-dean spaces. There is a point of contact with the embedding approach to gravity \cite{RegTeit,DesPirRob,PasFran}. Namely, if $f^A_\lambda = \partial_\lambda f^A$ for some $f^A$, the theory would be the theory of a four-dimensional hypersurface defined by the coordinates $f^A (x )$ in a ten-dimensional pseudo-Euclidean space-time. However, in general, $f^A_\lambda$ are independent freely chosen fields, and the Faddeev gravity can NOT be considered as an embedded theory of gravity.

The action does not contain squares of derivatives and, therefore, is finite for discontinuous $f^A_\lambda$ and $g_{\lambda \mu}$. This means that in quantum theory there are virtual configurations in which different regions do not fit on their common boundary due to the discontinuity of the induced metric on the boundary. That is, the metrics/fields $f$ in these regions can vary independently, though, of course, they can interact with each other, this interaction is simply not infinite, but finite.

In particular, a 2D surface can consist of virtually independent pieces, and its area spectrum is the sum of the spectra of the pieces. The area spectrum is important in the black hole physics.

For the area spectrum, it is natural to use a connection representation, in which tetrad bilinears are canonically conjugated to connection variables. It is fortunate that a connection Palatini-type form exists for the Faddeev gravity, as we have found in \cite{Kha1}. This representation is naturally obtained even for a more general form of the Faddeev action with a parity violating term,
\begin{equation}\label{Faddeev-gamma}                                       
\int \Pi^{AB} \left [ (f^\lambda_{A, \lambda} f^\mu_{B, \mu} - f^\lambda_{A, \mu} f^\mu_{B, \lambda}) \sqrt {g} - \frac{1}{\gamma_{\rm F}} \epsilon^{\lambda \mu \nu \rho} f_{\lambda A, \mu} f_{\nu B, \rho} \right ] \d^4 x.
\end{equation}

\noindent On the equations of motion, this still gives the Hilbert-Einstein action. Here $\gamma_{\rm F}$ is an analog of the Barbero-Immirzi parameter $\gamma$ \cite{Barb,Imm}. The Barbero-Immirzi parameter defines a term which generalizes the Cartan-Weyl form and does not affect the result of excluding the connection via equations of motion \cite{Holst,Fat}. The connection form is
\begin{eqnarray}\label{S Imm full}                                          
& & \hspace{-0mm} S_{\rm continuum} = \int \left ( f^\lambda_A f^\mu_B + \frac{1}{2 \gamma_{\rm F} \sqrt{g}} \epsilon^{\lambda \mu \nu \rho} f_{\nu A} f_{\rho B} \right ) \left [ \partial_\lambda \omega^{AB}_\mu - \partial_\mu \omega^{AB}_\lambda \right. \nonumber \\ & & \hspace{-5mm} \left. + (\omega_\lambda \omega_\mu - \omega_\mu \omega_\lambda)^{AB} \right ] (\omega ) \sqrt{g} \d^4 x + \int \Lambda^\nu_{\lambda \mu } \omega_{\nu}^{AB} \left ( f^\lambda_A f^\mu_B - f^\mu_A f^\lambda_B \right ) \sqrt{g} \d^4 x.
\end{eqnarray}

\noindent The first term is the so(10) connection Cartan-Weyl form of the Einstein-Hilbert action, the second term is an so(10) gauge violating term (Faddeev action is invariant with respect to the global, but not the local SO(10) rotations), $\Lambda^\nu_{\lambda \mu }$ are Lagrange multipliers at it.

Not tetrad bilinears, but area tensors have more relevance to the area. Tetrad bilinears go to area tensors in the minisuperspace formulation of gravity by Regge \cite{Regge} (see also review \cite{RegWil}). This minisuperspace formulation, Regge calculus, is automatically discrete.

We were able to find the Faddeev action for $f^A_\lambda$ and $g_{\lambda \mu}$ which are piecewise constant \cite{Kha} and, thus, discontinuous, however, with an uncertainty, which, however, disappears when the lattice spacings and field variations/dis\-continuities are made arbitrarily small, and we thus approach the continuum limit. This is a discrete Faddeev action, the Regge calculus form of the Faddeev action.

The notations are as follows. The piecewise flat spacetime can be viewed as composed of flat 4-dimensional tetrahedra or 4-simplices, although here most manipulations are made with a combinatorially simpler subdivision into hypercubes. The continuum field $f^A_\lambda$ is replaced by the variables $f^A_{\sigma^1} = f^A_\lambda \Delta x^\lambda_{\sigma^1}$ on the edges $\sigma^1$, where on the hypercubic structure the index $\sigma^1$ is replaced by its number, one of four at each vertex, and $\sigma^1$ formally looks like the world vector index. The connection becomes an SO(10) matrix $\Omega^A_{\sigma^3 B}$ on the tetrahedra (3-simplices) $\sigma^3$ or, on the hypercubic structure, $\Omega^A_{\lambda B}$ on the 3-cube complementary to the coordinate $x^\lambda$ (the coordinates of the vertices are assumed to run through integer four coordinates).

Then we can raise the question about a connection representation of the discrete Faddeev action with independent variables $\Omega^A_{\sigma^3 B}$, analogous to (\ref{S Imm full}) for the continuum Faddeev action. Such a representation exists and for the combinatorially simplified hypercubic structure just considered has the form \cite{Kha2}
\begin{eqnarray}\label{S-discr}                                             
& & S = S_{R (\Omega )} + S_\Omega, ~~~ S_{R (\Omega )} = 2 \sum_{\rm sites} \sum_{\lambda, \mu} \left[ a^{\lambda \mu} \arcsin \frac{v^{\lambda \mu} \circ R_{\lambda \mu}(\Omega ) }{a^{\lambda \mu} } \right. \nonumber \\ & & \hspace{-7mm}  \left. + \frac{a^{\lambda \mu}}{\gamma_{\rm F}} \arcsin \frac{V^{\lambda \mu} \circ R_{\lambda \mu}(\Omega ) }{a^{\lambda \mu} } \right], ~~~ S_\Omega = \sum_{\rm sites} \sum_{\lambda, \mu, \nu} \Lambda^\nu_{\lambda \mu} \Omega^{AB}_\nu (f^\lambda_A f^\mu_B - f^\mu_A f^\lambda_B),
\end{eqnarray}

\noindent $a^{\lambda \mu} = \sqrt{v^{\lambda \mu} \circ v^{\lambda \mu} } = \sqrt{V^{\lambda \mu} \circ V^{\lambda \mu} }$, $R_{\lambda\mu} (\Omega ) = \overOm_\lambda (\overT_\lambda \overOm_\mu) (\overT_\mu \Omega_\lambda) \Omega_\mu $. The area bivectors are
\begin{equation}\label{vV}                                                  
v^{\lambda \mu}_{A B} = \frac{1}{2} \left ( f^\lambda_A f^\mu_B - f^\mu_A f^\lambda_B \right ) \sqrt{g}, ~~~ V^{\lambda \mu}_{A B} = \frac{1}{2} \epsilon^{\lambda \mu \nu \rho} f_{\nu A} f_{\rho B}, ~~~ v \circ R \equiv \frac{1}{2} v_{AB} R^{AB},
\end{equation}

\noindent $T_\lambda f (x^\lambda ) = f (x^\lambda + 1)$, the operator of the translation along $x^\lambda$ by 1, the overlining in $\overOm$, $\overT$, ... means the Hermitian conjugation. Since the vertical-vertical block of the {\it continual} $\omega$ does not contribute to $S_{\rm continuum}$, we can impose also a condition $\Pi_{A C} \Pi_{B D} \Omega^{C D}_\lambda = 0$ (now $\Omega = \exp \omega$), which does not change the continuum limit.

The first term in (\ref{S-discr}) is the SO(10) connection representation of the Regge action, the second term is a discretization of the gauge violating term in (\ref{S Imm full}).

As before for the discrete Faddeev action, (\ref{S-discr}) has the exact sense of the connection representation of the Faddeev action just in the continuum limit, when $\omega$ ($\Omega = \exp \omega$) and variations $\delta f$ from point to point are arbitrarily small.

In the Regge-discretized connection representation of the Faddeev action (\ref{S-discr}), area tensors are canonically conjugate to SO(10) connection matrices, and their spectrum can be found \cite{kha-spectrum}.

A certain subtlety is that the kinetic term ${\rm tr} ( A^{0 \lambda} \overOm_\lambda \dotOm_\lambda ) $, $A^{\lambda \mu} = v^{\lambda \mu} + V^{\lambda \mu} / \gamma_{\rm F}$, due to the gauge violating condition on $\Omega = \exp \omega$ ($\omega$ has a zero horizontal-horizontal block) is nonzero only starting from the second order in $\omega$, ${\rm tr} ( A^{0 \lambda} \omega_\lambda \dotom_\lambda ) $.

Then for small $\omega$ the area spectrum is singular.

The possibility to get around this difficulty is that the metric can vary weakly also for strongly varying $f^A_\lambda$ and thus for large $\omega$.

Thus, the task is to extend the Regge-discretized connection representation of the Faddeev action (\ref{S-discr}) to large connections.

To generalize the action to the large $\omega$, we can start with the case when only one component $\omega_\lambda$ is large and use the fact that the equation for $\omega_\lambda$ follows by varying over $\omega_\mu$, $\mu \neq \lambda$, which are small and the dependence on them is known. It is also natural to suggest that the large $\omega_\lambda$ corresponds to a strong variation of the tetrad as a function of $x^\lambda$, as qualitatively follows by extrapolation from the case of small $\omega$. Having found the dependence on the large connection matrix, we can consider the case when all the connection matrices are large. For clarity, we use a hypercubic lattice; then we pass to the general simplicial complex.

\section{Generalizing local SO(10) violating term to large connection}

Here $S_\Omega$ is generalized to not necessarily small connection in a certain direction, (\ref{lambda3term}). Due to a specific form of the equations for the connection and natural requirements of correspondence with the solutions at small variations of $f^\lambda_A$, the form of this term follows practically uniquely.

We denote the relative order of magnitude of the variations of the metric $\delta g$ from point to point by $O( \delta )$ (and, under the assumption of smoothness in the continuum limit, $O( \delta^2 ) = O( \delta )^2$). Normally, the variations of the Faddeev tetrad $\delta f$ from point to point have the same relative order. Based on the answer, $\omega = O( \delta )$, $r = O( \delta^2 )$ ($\Omega = \exp \omega$, $R = \exp r$). Varying over $\Omega$, we have for $S_{R (\Omega )}$ in the order $O( \delta )$
\begin{eqnarray}\label{dS/domega}                                           
& & \hspace{-8mm} M^\mu_{A B} S_{R (\Omega )} = 2\sum_\lambda [ A^{\lambda \mu} - T_\lambda (\Omega_\lambda A^{\lambda \mu} \overOm_\lambda)], ~~~ M^\mu_{A B} = \Omega^{C}_{\mu A} \frac {\partial } {\partial \Omega^{CB}_\mu } - ( A \leftrightarrow B ), \nonumber \\ & & A^{\lambda \mu} \equiv v^{\lambda \mu} + \frac{1}{\gamma_{\rm F}} V^{\lambda \mu},
\end{eqnarray}

\noindent and for $S_\Omega$
\begin{equation}\label{dgauge-viol/domega}                                  
M^\mu_{A B} S_\Omega = \sum_{\nu, \lambda} \Lambda^\mu_{\nu \lambda} \{ \overOm_\mu [f^\nu , f^\lambda] + [f^\nu , f^\lambda] \Omega_\mu \}_{A B}, ~~~ [f^\nu , f^\lambda] \equiv f^\nu_A f^\lambda_B - f^\lambda_A f^\nu_B .
\end{equation}

\noindent The operator $M^\mu_{A B}$ cancels the orthogonality condition on $\Omega_\mu$ added to $S$ with the help of Lagrange multipliers not shown here. Equating the sum of (\ref{dS/domega}) and (\ref{dgauge-viol/domega}) we apply first the projector $\Pi_\| = 1 - \Pi$ on both sides, $\Pi_\| \dots \Pi_\|$, find that the six components $\Lambda^\mu_{\nu \lambda}$ (for the given $\mu$) are $O ( \delta )$, then apply $\Pi \dots \Pi_\|$ or simply $\Pi $ and find that (\ref{dgauge-viol/domega}) contributes $O ( \delta^2 )$ and can be disregarded. Thus, in the order $O ( \delta )$ we have
\begin{equation}\label{omega-eq}                                           
\Pi \sum_\lambda T_\lambda \{ \delta_\lambda A^{\lambda \mu} + [\omega_\lambda, A^{\lambda \mu} ] \} = 0, ~~~ \delta_\lambda = 1 - \overT_\lambda.
\end{equation}

\noindent This gives a solution
\begin{equation}\label{omega}                                              
\omega_\lambda = [ f^\mu , \Pi \delta_\lambda f_\mu ].
\end{equation}

\noindent Thus, $\omega_\lambda$ is defined by the dependence of $f$ on $x^\lambda$.

First assume that there is a strong variation only in one direction, say, along the coordinate $x^3$. It is natural to expect that this will not affect, at least in the considered order $O ( \delta )$, the expressions for $\omega_\lambda$, $\lambda \neq 3$, which remain small, and the terms in $S_\Omega$ for them remain the same. And the equations for $\Omega_3$ are obtained by varying just over $\Omega_\mu$, $\mu \neq 3$. Assuming still $r = O ( \delta^2 )$ ($R = \exp r$) and that the result of the action of the operator $M^\mu_{A B}$ on $S_{R (\Omega )}$ is $O ( \delta )$ (as confirmed by the further calculation), this result is again (\ref{dS/domega}) and $\Lambda^\mu_{\nu \lambda} = O ( \delta )$, $\mu \neq 3$, and the equation for $\Omega$ is again (\ref{dS/domega}) projected by $\Pi$ and equated to zero. The expression (\ref{dS/domega}) is the sum of the expression
\begin{equation}\label{Omega3eq}                                           
A^{3 \mu} - T_3 (\Omega_3 A^{3 \mu} \overOm_3)
\end{equation}

\noindent and terms with $\omega_\lambda$ at $\lambda \neq 3, \mu$, which being projected by $\Pi$ are $O ( \delta^2 )$ and can be disregarded. Thus, the question is about (\ref{Omega3eq}) whether its projection by $\Pi$ can be zero in the order $O ( \delta )$. The form of (\ref{Omega3eq}) suggests that for a tetrad whose values at neighboring points are roughly related by an orthogonal rotation, one can make a redefinition of the tetrad and the connection at these points by this rotation. Showing the dependence of the functions on $x^3$, we introduce for generality such an orthogonal matrix at each point and the notation
\begin{equation}                                                           
\O ( 0 ) f^\lambda ( 0 ) = h^\lambda ( 0 ), ~~~ \O (-1 ) f^\lambda (-1 ) = h^\lambda (-1 )
\end{equation}

\noindent for the tetrad $f^\lambda$ at two neighboring points $x^3 = -1$ and $x^3 = 0$, the orthogonal matrices $\O ( -1 ), \O ( 0 ) \in $ SO(10) at these points and an interpolating field $h ( x^3 ) $, whose variations $\delta_3 h^\lambda ( 0 ) = h^\lambda ( 0 ) - h^\lambda ( -1 )$ are small. Correspondingly, we write out
\begin{equation}                                                           
\Omega_3 (0) = \overO (-1) [1 + \omega_3 (0)] \O (0).
\end{equation}

\noindent Then, for example,
\begin{eqnarray}                                                           
& & \Pi (0) \equiv \Pi ( f (0)) = \overO (0) \Pi (h(0)) \O (0), ~ A^{3 \mu } (0) \equiv A^{3 \mu } ( f (0)) \nonumber \\ & & = \overO (0) A^{3 \mu } (h(0)) \O (0)
\end{eqnarray}

\noindent and
\begin{eqnarray}                                                           
& & \Pi (-1) \{ A^{3 \mu} (-1) - T_3 [\Omega_3 A^{3 \mu} \overO_3]_{x^3 = -1} \} \nonumber \\ & & = \Pi (-1) [ A^{3 \mu} (-1) - \Omega_3 (0) A^{3 \mu} (0) \overO_3 (0) ] \nonumber \\& &  = - \overO (-1) \Pi (h (-1)) \{ \delta_3 A^{3 \mu} (h(0)) + [\omega_3 (0), A^{3 \mu} (h(0)) ] \} \O (-1).
\end{eqnarray}

\noindent This is zero in the order $O ( \delta )$ at
\begin{equation}\label{omega3}                                             
\omega_3 (0) = [h^\lambda , \Pi \delta_3 h_\lambda ] (0).
\end{equation}

We have used the known terms in $S_\Omega$ for $\omega_\lambda, \lambda \neq 3$, and not for $\omega_3$. Now we can restore such a term corresponding to the found $\omega_3$, vary over $\omega_3$ and check that the corresponding equation of motion is satisfied as well. The term in $S_\Omega$ is readily read from the expression for $\omega_3$ (\ref{omega3}),
\begin{equation}\label{lambda3term}                                        
\Lambda^3_{\nu \lambda} (1 + \omega_3 (0))^{AB} [h^\nu , h^\lambda ]_{AB} (0) = \Lambda^3_{\nu \lambda} (\overU_3 \Omega_3)^{A B} (0) [ f^\nu , f^\lambda ]_{A B} (0).
\end{equation}

\noindent Here $U_3 (0) \equiv \overO (-1) \O (0)$. The action of the operator $M^\mu_{A B}$ on this term leads to
\begin{equation}\label{dgauge-viol/domega3}                                
\Lambda^3_{\nu \lambda} \{ [1 - \overO (0) \omega_3 (0)) \O (0)] [ f^\nu , f^\lambda ] + [ f^\nu , f^\lambda ] [1 + \overO (0) \omega_3 (0)) \O (0)] \}_{AB}.
\end{equation}

\noindent Applying $M^\mu_{A B}$ to the total $S^{\rm discr}$ and equating this to zero, we get the equation for $\omega_3$ (in fact, a condition on $\omega_\lambda$, $\lambda \neq 3$) and $\Lambda^3_{\nu \lambda}$. As before, projecting this by $\Pi_\| \dots \Pi_\|$ we get $\Lambda^3_{\nu \lambda} = O ( \delta )$, then the contribution of the term in $S_\Omega$ (\ref{dgauge-viol/domega3}) can be neglected upon projecting by $\Pi$. The equation becomes (\ref{omega-eq}) at $\mu = 3$, and this is satisfied in the order $O ( \delta )$ by the solution (\ref{omega}).

\section{The tetrad fields strongly varying in all directions}

Here we arrive at $S_\Omega$ of the form (\ref{U-omega-ff}) with $U_\lambda$ minimizing expressions of the type of (\ref{f-Uf}) and $S$ on the general simplicial complex (\ref{S-discr-general}), which, depending on $U$, can be viewed as a unified representation of the Faddeev or Regge action.

As before, since Faddeev gravity action is ambiguous on the piecewise flat space-time (depends on the intermediate regularization of the conical singularities) and the ambiguity decreases to zero as we approach the continuum limit, we adopt the behavior of the variables corresponding to approaching a continuum metric field. Namely, the variations of the metric tensor $\delta_\nu (f^\lambda_A f^{\mu A} )$ are small (the order of such a smallness is just denoted as $O (\delta )$). Then the lengths of the tetrad vectors $f^\lambda_A$ and the angles between them at the vertices $x$ and $x - \delta_\mu x$ differ by $O (\delta )$, so if we choose a matrix $U_\mu (x) \in $ SO(10) that minimizes $f^\lambda (x - \delta_\mu ) - U_\mu (x) f^\lambda (x)$ in one sense or another, this quantity will be of the order of $O (\delta )$. (Here the components of $\delta_\mu x$ are literally $(\delta_\mu x)^\lambda = [(1 - \overT_\mu ) x ]^\lambda = \delta^\lambda_\mu$.)
\begin{equation}\label{f-Uf}                                               
f^\lambda (x - \delta_\mu x ) - U_\mu (x) f^\lambda (x) = O (\delta ).
\end{equation}

\noindent Of course, it is scalar (with respect to the local SO(10)) quantities like $f^\nu_A (x - \delta_\mu ) [f^\lambda (x - \delta_\mu ) - U_\mu (x) f^\lambda (x)]^A$ which should be minimized, therefore it follows that SO(10) rotations of the tetrad at $x$ and at $x - \delta_\mu x$ induce the transformation of $U_\mu (x)$ as a (finite) connection. Relating in this way $f^\lambda (x - \delta_\mu x - \delta_\nu x )$ to $f^\lambda (x )$ through, first, $f^\lambda (x - \delta_\mu x )$ and, second, $f^\lambda (x - \delta_\nu x )$ and comparing, we find for the holonomy of this connection
\begin{equation}                                                           
\overU_\nu (x) \overU_\mu (x - \delta_\nu x ) U_\nu (x - \delta_\mu x ) U_\mu (x) = 1 + O (\delta ).
\end{equation}

\noindent In the leading order in $\delta$, this holonomy is 1 and is solved by the SO(10) rotations in the hypercubes so that
\begin{equation}\label{U=OO}                                               
U_\lambda (x) = \overO (x - \delta_\lambda x) \O (x)
\end{equation}

\noindent on the 3-face between the hypercubes at $x$ and at $x - \delta_\lambda x$.

Then we can describe the system as including the following $S_\Omega$,
\begin{equation}\label{U-omega-ff}                                         
S_\Omega = \sum_x \sum_{\lambda, \mu, \nu} \Lambda^\mu_{\nu \lambda} [\overU_\mu (x) \Omega_\mu (x) ]^{AB} [f^\nu (x), f^\lambda (x)]_{AB}.
\end{equation}

\noindent The solution for $\Omega$ can be written in the form
\begin{equation}                                                           
\Omega_\lambda (x) = \overO (x - \delta_\lambda x) [ 1 + \omega_\lambda (x) ] \O (x)
\end{equation}

\noindent where $\omega_\lambda (x)$ in terms of the interpolating field
\begin{equation}\label{h=Of}                                               
h^\lambda (x) = \O (x) f^\lambda (x)
\end{equation}

\noindent follows by replacing $f$ in (\ref{omega}) by $h$,
\begin{equation}                                                           
\omega_\lambda (x) = [ h^\lambda (x), \Pi (h(x)) \delta_\lambda h_\mu (x) ].
\end{equation}

Indeed, the result of action of the operator $M^\mu_{A B}$ on $S_{R (\Omega )}$ (\ref{dS/domega}) can be rewritten as
\begin{eqnarray}                                                           
& & 2 \sum_\lambda T_\lambda [A^{\lambda \mu} (f(x - \delta_\lambda x)) - \Omega_\lambda (x) A^{\lambda \mu} (f(x)) \overOm_\lambda (x)] = 2 \sum_\lambda T_\lambda \overO (x - \delta_\lambda x ) \nonumber \\ & & \cdot [A^{\lambda \mu} (h(x - \delta_\lambda x)) - (1 + \omega_\lambda (x) ) A^{\lambda \mu} (h(x)) (1 - \omega_\lambda (x) ) ] \O (x - \delta_\lambda x) \nonumber \\ & & = 2 \sum_\lambda \overO (x) \{ T_\lambda [ - \delta_\lambda A^{\lambda \mu} (h(x)) - [ \omega_\lambda (x), A^{\lambda \mu} (h(x))] ] \} \O (x),
\end{eqnarray}

\noindent the result of such an action on $S_\Omega$ is
\begin{equation}                                                           
\sum_{\nu, \lambda} \Lambda^\mu_{\nu, \lambda} \overO (x) \{ (1 - \omega_\lambda (x) ) [h^\nu , h^\lambda ] + [h^\nu , h^\lambda ] (1 + \omega_\lambda (x) ) \} \O (x).
\end{equation}

\noindent The resulting equations for $\Omega$ follow from those for $f$ at small $\delta f$ by replacing $f^\lambda_A$ by $h^\lambda_A$. Correspondingly, the Faddeev action following upon excluding $\Omega$ will be (\ref{Faddeev-gamma}) with $f^\lambda_A$ replaced by $h^\lambda_A$. Since $h^\lambda_A h^{\mu A} = f^\lambda_A f^{\mu A} = g^{\lambda \mu}$, this will be equivalent to general relativity with the metric $g^{\lambda \mu}$.

Thus, the connection representation of the general Faddeev action with strongly varying fields is (\ref{S-discr}) with $S_\Omega$ generalized to (\ref{U-omega-ff}).

This procedure seems to be sufficiently algorithmized to introduce it for the true mini-superspace of the general simplicial complex. Instead of (\ref{f-Uf}), we have
\begin{equation}\label{f-Ufsigma}                                          
f_{\sigma^1_i | \sigma^4_2} - U_{\sigma^3}f_{\sigma^1_i | \sigma^4_1} = O (\delta )
\end{equation}

\noindent for some four edges $\sigma^1_i$, $i = 0, 1, 2, 3$, of one of any two 4-simplices $\sigma^4_1$ and $\sigma^4_2$ having a common 3-face $\sigma^3 = \sigma^4_1 \cap \sigma^4_2$; $f_{\sigma^1 | \sigma^4}$ means that the vector $f^A_{\sigma^1}$ is defined in the frame associated with the 4-simplex $\sigma^4$. In the leading order in $\delta$, this has the solution with the trivial holonomy,
\begin{equation}\label{U=OOsigma}                                          
U_{\sigma^3} = \overO_{\sigma^4_2} \O_{\sigma^4_1}.
\end{equation}

On the general simplicial complex, the connection action takes the form
\begin{eqnarray}\label{S-discr-general}                                    
& & S = 2 \sum_{\sigma^2} \left[ a_{\sigma^2} \arcsin \frac{v_{\sigma^2} \circ R_{\sigma^2}(\Omega ) }{a_{\sigma^2} } + \frac{a_{\sigma^2}}{\gamma_{\rm F}} \arcsin \frac{V_{\sigma^2} \circ R_{\sigma^2}(\Omega ) }{a_{\sigma^2} } \right] \nonumber \\ & & \hspace{-8mm} + \sum_{\sigma^3} \sum_{\{ \sigma^2 : ~ \sigma^2 \subset \sigma^4 (\sigma^3 ) \} } \Lambda^{\sigma^3 \sigma^2} (\overU_{\sigma^3} \Omega_{\sigma^3})_{A B} V_{\sigma^2 | \sigma^4 (\sigma^3)}^{AB}, ~ R_{\sigma^2} = \prod_{\sigma^3 \supset \sigma^2} \Omega^{\epsilon (\sigma^2, \sigma^3)}_{\sigma^3},
\end{eqnarray}

\noindent $a_{\sigma^2} = \sqrt{v_{\sigma^2} \circ v_{\sigma^2} } = \sqrt{V_{\sigma^2} \circ V_{\sigma^2} }$, $\sigma^4 (\sigma^3 )$ is one of the two 4-simplices sharing $\sigma^3$, $\epsilon (\sigma^2, \sigma^3) = \pm 1$ is some sign function. Here the bivector of the triangle $\sigma^2$ and the dual one in terms of the edge vectors are
\begin{equation}\label{Vv}                                                 
V^{AB}_{\sigma^2} = \frac{1}{2} ( f^A_{\sigma^1_1} f^B_{\sigma^1_2} - f^B_{\sigma^1_1} f^A_{\sigma^1_2} ), ~~~ v_{\sigma^2 AB} = \frac{1}{2} \epsilon_{ABCD} (\sigma^4 ) V^{CD}_{\sigma^2},
\end{equation}

\noindent where the bivectors are defined in the local frame of a 4-simplex $\sigma^4 \supset \sigma^2$ and
\begin{equation}                                                           
\epsilon_{ABCD} (\sigma^4 ) = \frac{\epsilon^{\tson_1 \tson_2 \tson_3 \tson_4} f_{\tson_1 A} f_{\tson_2 B} f_{\tson_3 C} f_{\tson_4 D}}{\sqrt{\det \| f_{\tson_1 A} f^A_{\tson_2} \|}}
\end{equation}

\noindent is the perfectly antisymmetric fourth rank tensor in the "horizontal" 4-dimensional subspace. Here, $\epsilon^{\tson_1 \tson_2 \tson_3 \tson_4} = \pm 1$ is the parity of a permutation $(\tson_1 \tson_2 \tson_3 \tson_4 )$ of a quadruple of edges $(\son_1 \son_2 \son_3 \son_4 )$ which span the given 4-simplex. The sum over all the permutations is implied.

In (\ref{S-discr-general}), $U_{\sigma^3}$ serves as an auxiliary connection but it is a (somewhat loosely fixed) function of the tetrad (edge vector) variables and refers to the tetrad sector of variables. The above basic choice of it leading to the Faddeev gravity is a connection for the system in the regime of approaching the continuum limit in the leading order (and thus with the trivial holonomy). If $U_{\sigma^3}$ are exact rotations relating neighboring frames, then (\ref{S-discr-general}) is a representation of the Regge action. Indeed, varying $S$ over $\Omega_{\sigma^3}$ at $\Omega_{\tilde{\sigma}^3} = U_{\tilde{\sigma}^3} \forall \tilde{\sigma}^3$, we get just the sum of the area tensors over the surface of the tetrahedron $\sigma^3$, which is zero, and a linear in $\Lambda$ term. Together with $\Lambda = 0$, this solves the equations for $\Omega$ and $\partial S / \partial \Lambda = 0$, and we have the Regge action.

\section{Conclusions}

1) We have considered the case of a variation of the Faddeev fields, when $\Omega_{\sigma^3}$ for all orientations of the 3-face $\sigma^3$ can differ substantially from unity. Or, in the particular case of the hypercubic structure, $\Omega_\lambda$ for all $\lambda$ can significantly differ from unity. This, in particular, makes it possible to have a reasonable spectrum for elementary areas of any orientation and, simultaneously, general relativity on a large scale on classical level. In the case of the hypercubic structure, we have the local SO(10) violating term (\ref{U-omega-ff}) in the connection representation of the type of (\ref{S-discr}) with $U_\lambda$ minimizing expressions of the type of (\ref{f-Uf}). Excluding connection gives the generalized Faddeev action (\ref{Faddeev-gamma}) with $f^\lambda_A$ replaced by $( \O f^\lambda )_A$ (\ref{h=Of}) with SO(10) rotations in the hypercubes $\O$ solving for $U_\lambda$ according to (\ref{U=OO}) in the continuum limit.

2) The discrete connection representation for the Faddeev gravity can be considered as a particular case of the unified representation (\ref{S-discr-general}) for both the Regge and discrete Faddeev formulations depending on the auxiliary SO(10) field variable $U_{\sigma^3}$ minimizing expressions of the type of (\ref{f-Ufsigma}) and referred to the tetrad sector of the theory (a certain function of the tetrad or edge vectors) but being a certain connection by its local frame transformation properties. If the system is in the regime of approaching the continuum limit when the metric field can arbitrarily weakly vary from vertex to vertex and $U_{\sigma^3}$ is taken in the leading order (and thus with the trivial holonomy), we get the representation of the Faddeev gravity. If we set $U_{\sigma^3}$ to be exact rotations relating neighboring frames, it is a representation of the Regge action.

The present analysis does not contain the examination of our previous work \cite{kha-strongly} as a particular case: we consider there the connection that has a large leading part $U_\lambda$ only for a certain $\lambda$, but $U_\lambda$ depends on $x$ and has a nontrivial holonomy ($U$ is not written in terms of $\O$ as in (\ref{U=OO}) or (\ref{U=OOsigma}). As a result, we obtained in that paper in some respects a generalization of the Faddeev gravity (in particular, in the continuum limit, a certain sum of coupled systems of the Faddeev type), although in another respect a less general system with a strong variation of $f^\lambda_A$ in only one direction.

\section*{Acknowledgments}

The present work was supported by the Ministry of Education and Science of the Russian Federation

\end{document}